\documentclass{ws-procs9x6}

\usepackage{graphicx}     

\begin{document}

\title{SUPERFLUID-NORMAL PHASE TRANSITION \\ IN FINITE SYSTEMS
AND ITS EFFECT ON DAMPING\\ OF HOT GIANT RESONANCES\footnote{Invited
lecture at the Predeal international summer school in nuclear physics
on ``Collective motion and phase transitions in nuclear systems'', 28 
August - 9 September, 2006, Predeal, Romania}}
\author{NGUYEN DINH DANG$^\dag$}
\address{
    1) Heavy-ion nuclear physics laboratory, Nishina Center for
Accelerator-Based Science, RIKEN, 2-1 Hirosawa, Wako city, 351-0198 
 Saitama, Japan\\
2) Institute for Nuclear Science and Technique, Hanoi, Vietnam
\\$^\dag$E-mail: dang@riken.jp}
\begin{abstract}
   Thermal fluctuations of quasiparticle number are included making
   use of the secondary Bogolyubov's transformation, which turns
   quasiparticles operators into modified-quasiparticle ones. This
   restores the unitarity relation for the generalized single-particle
   density operator, which is violated within the
   Hartree-Fock-Bogolyubov (HFB) theory at finite temperature. The resulting
   theory is called the modified HFB (MHFB) theory, whose limit of a
   constant pairing interaction yields the modified BCS (MBCS) theory.
   Within the MBCS theory, the pairing gap never collapses at finite temperature 
   $T$ as it does within the BCS theory, but decreases monotonously with increasing
   $T$. It is demonstrated that this non-vanishing thermal
   pairing is the reason why the width of the giant dipole
   resonance (GDR) does not increase with $T$ up to $T\sim$ 1 MeV. At 
   higher $T$, when the thermal pairing is small, the
   GDR width starts to increase with $T$. The calculations within the 
   phonon-damping model yield the results in good agreement with the
   most recent experimental systematic for the GDR width as a function
   of $T$. A similar effect, which causes a small GDR width at low $T$, 
   is also seen after thermal pairing is included in the thermal
   fluctuation model.
   
\end{abstract}

\bodymatter

\section{Introduction}
It is well known that infinite systems undergo a sharp phase transition from the superfluid 
phase to the normal-fluid one at finite temperature $T$. 
Marked by a
collapse of the pairing correlations (pairing gap), and a
near divergence of the heat capacity at a critical temperature
$T_{c}$, this phase transition is a second-order one.  The critical temperature is
found to be $T_{c}\simeq$ 0.567$\Delta(0)$ for infinite systems,
where $\Delta(0)$ is the pairing gap at zero temperature $T=$
0~\cite{Landau}.

The application of the BCS theory and its generalization, the
Hartree-Fock-Bogolyubov
(HFB) theory, to finite Fermi systems paved the 
way to study the superfluid-normal (SN) phase transition in nuclei at
finite temperature~\cite{Sano,Goodman1,Goodman2}. Soon it has been realized that the BCS
and HFB theories ignore a number of quantal and thermodynamic
fluctuations, which become large in small systems because of their
finiteness. 
As a consequence, the unitarity
relation for the generalized particle-density  matrix $R$, which
requires $R^{2}=R$, is violated. In deed, within the HFB 
theory at $T\neq$ 0, one has $Tr[R^{2}(T)-R(T)]=2\delta{\cal N}^2\equiv2\sum_{i}\delta{\cal
    N}_{i}^{2}=2\sum_{i}n_{i}(1-n_{i})>$ 0
where $n_{i}=[{\rm e}^{\beta E_{i}}+1]^{-1}$
is the occupation number of non-interacting quasiparticles with energy
$E_{i}$ at temperature $T=1/\beta$ on the $i$-th
orbital~\cite{Goodman2}. 
Large thermal fluctuations smooth out the sharp second-order SN phase transition. 
As the result the pairing gap does not collapse as has been predicted by 
the BCS theory, but decreases monotonously as the temperature
increases, and remains finite even at rather high
$T$~\cite{Moretto,DangZ,DangRing}.
So far these fluctuations were taken into account based on the macroscopic Landau theory 
of phase transitions~\cite{Moretto,DangZ}. 
This concept is close to that of the static-path
approximation, which treats thermal fluctuations on all possible
static paths around the mean field~\cite{DangRing}.

It will be shown in the first part of this lecture that 
the recently proposed modified-BCS (MBCS) theory~\cite{MBCS1,MBCS}, 
and its generalization, the modified-HFB (MHFB)
theory~\cite{MHFB} take into
account the fluctuations of
quasiparticle number in a microscopic way. The MHFB
theory 
restores the unitarity relation by explicitly including the quasiparticle-number
fluctuations, making use of a
secondary Bogolyubov transformation from quasiparticle operators to
modified quasiparticle ones. 
In the limiting case of a constant
pairing interaction $G$ the MHFB equation is reduced to the MBCS one.

The second part of the lecture represents an application of the MBCS theory in 
the study of the damping of giant dipole resonances (GDR) in
hot nuclei, which are formed at high excitation energies $E^{*}$ in heavy-ion fusion reactions
or in the inelastic scattering of light particles (nuclei) on heavy 
targets. The $\gamma$-decay spectra of these compound nuclei show the existence of the
GDR, whose peak's energy depends weakly on the excitation energy $E^{*}$.
The dependence of the GDR on the temperature $T$ has been experimentally
extracted when the angular momentum of the compound nucleus is low, as in the case 
of the light-particle scattering experiments, or when it can be separated out
from the excitation energy $E^{*}$. These measurements have showed that the GDR 
width remains almost constant at
$T\leq$ 1 MeV, but sharply increases with $T$ up to $T\simeq$ 2 - 3 MeV,
and saturates at higher $T$~\cite{Woude}. The phonon-damping model
(PDM), proposed by the lecturer in collaboration with
Arima~\cite{PDM}, explains the GDR width's increase and saturation by 
coupling the GDR to non-collective particle-particle ($pp$) and
hole-hole ($hh$) configurations, which appear  due to the 
deformation of the Fermi surface at $T\neq$ 0.
It will be shown that, by including non-vanishing MBCS thermal pairing, the 
PDM is
also able to predict the GDR width at low $T$.

\section{Modified HFB theory at finite temperature and its limit, modified BCS theory}
\subsection{HFB theory}
The HFB theory is based on the self-consistent Hartree-Fock (HF) 
Hamiltonian with two-body interaction
\begin{equation}
H=\sum_{ij}{\cal T}_{ij}a_{i}^{\dagger}a_{j}+\frac{1}{4}\sum_{ijkl}
v_{ijkl}a_{i}^{\dagger}a_{j}^{\dagger}a_{l}a_{k}~,
\label{HHF}
\end{equation}
where $i,j,..$ denote the quantum numbers 
characterizing the single-particle 
orbitals, ${\cal T}_{ij}$ are the 
kinetic energies, and $v_{ijkl}$ are  
antisymmetrized matrix elements of the two-body interaction. 
The HFB theory approximates Hamiltonian (\ref{HHF}) by an 
independent-quasiparticle Hamiltonian $H_{\rm HFB}$
\begin{equation}
H-\mu\hat{N}\approx H_{\rm 
HFB}=E_{0}+\sum_{i}E_{i}\alpha_{i}^{\dagger}\alpha_{i}~,
\label{HHFB}
\end{equation}
where $\hat{N}$ is the particle-number operator, $\mu$ is the 
chemical 
potential, $E_{0}$ is the energy of the ground-state 
$|0\rangle$, which 
is defined as the vacuum of quasiparticles:
\begin{equation}
\alpha_{i}|0\rangle=0~,
\label{qpvacuum}
\end{equation}
and $E_{i}$ are quasiparticle energies. 
The quasiparticle creation
$\alpha_{i}^{\dagger}$ and destruction $\alpha_{i}$ operators are
obtained from the single-particle operators $a_{i}^{\dagger}$ and
$a_{i}$ by the Bogolyubov transformation,
whose matrix form is
\begin{equation}
	\left( \begin{array}{c}\alpha^{\dagger}\\
	\alpha\end{array}
	\right)=
	\left(\begin{array}{cc}
	U&V\\V^{*}&U^{*}
	\end{array}\right)
	\left(\begin{array}{c}
	a^{\dagger}\\ 
	a\end{array}\right)~
     \label{UV}
     \end{equation}
with the properties
\begin{equation}
UU^{\dagger}+VV^{\dagger}={\bf 1}~,\hspace{5mm} 
U{V}^{\rm T}+V{U}^{\rm T}=0~,
\label{UVproperties}
\end{equation}
where ${\bf 1}$ is the unit matrix, and the superscript $^{\rm T}$  
denotes the transposing operation. 
The quasiparticle energies $E_{i}$ and matrices $U$ and $V$ are 
determined 
as the solutions of the HFB equations, which are usually derived by 
applying either 
the variational principle of Ritz or the Wick's 
theorem. 

At finite temperature $T$ the condition for a system to be in 
thermal 
equilibrium requires the minimum of its grand potential $\Omega$
    \begin{equation}
    \Omega={\cal E}-TS-\mu N~,
    \label{Omega}
    \end{equation}
with the total energy ${\cal E}$, the entropy $S$, and particle 
number $N$, namely
\begin{equation}
    \delta\Omega=0~.
    \label{minimum}
    \end{equation}
    This variation defines the density operator ${\cal D}$ with
the trace equal to 1
 \begin{equation}
     {\rm Tr}{\cal D}=1~,\hspace{5mm} \delta\Omega/\delta{\cal D}=0~
     \label{dOmdD}
     \end{equation}
in the form
     \begin{equation}
	 {\cal D}=Z^{-1}{\rm e}^{-\beta(H-\mu\hat{N})},\hspace{5mm} 
	 Z={{\rm Tr}
	 [{\rm e}^{-\beta(H-\mu\hat{N})}]}~,\hspace{5mm} \beta=T^{-1}~,
	 \label{Dcal}
     \end{equation}
     where $Z$ is the grand partition function. The expectation value 
$\prec\hat{\cal O}\succ$ of 
     any operator $\hat{\cal O}$ is then given as the average in the 
grand canonical ensemble
     \begin{equation}
	 \prec\hat{\cal O}\succ = {\rm Tr}({\cal D}\hat{\cal O})~.
	 \label{average}
	 \end{equation}
	 This defines the total energy ${\cal E}$, entropy $S$, and 
	 particle number $N$ as 
     \begin{equation}
	 {\cal E}={\rm Tr}({\cal D}H)~,\hspace{3mm} 
	 S=-{\rm Tr}({\cal D}{\rm 
ln}{\cal D})~,
	 \hspace{3mm} N={\rm Tr}({\cal D}\hat{N})~.
	 \label{ESN}
	 \end{equation}	 
The FT-HFB theory replaces the unknown exact density operator 
${\cal D}$ in Eq. (\ref{Dcal}) 
with the approximated one, ${\cal D}_{\rm HFB}$, which is found in Ref. 
\cite{Goodman1} by substituting Eq. (\ref{HHFB}) in to Eq. 
(\ref{Dcal}) as
\begin{equation}
 {\cal D}_{\rm HFB}=
\prod_{i}[n_{i}\hat{\cal N}_{i}+(1-n_{i})
 (1-\hat{\cal N}_{i})]~,
 \label{DHFB}
 \end{equation}
where $\hat{\cal N}_{i}$ is the operator of quasiparticle number 
on the $i$-th orbital
     \begin{equation}
	 \hat{\cal N}_{i}=\alpha_{i}^{\dagger}\alpha_{i}~,
	 \label{Ncali}
	 \end{equation}
and $n_{i}$ is the quasiparticle occupation number. Within the FT-HFB 
theory $n_{i}$ is defined according to Eq. (\ref{average}) as
\begin{equation}
    n_{i}=\langle\hat{\cal N}_{i}\rangle= \frac{1}{{\rm e}^{\beta E_{i}}+1}~,
     \label{ni}
     \end{equation}
where the symbol $\langle\ldots\rangle$ denotes the 
average similar to (\ref{average}), but in which the approximated density operator 
${\cal D}_{\rm HFB}$ (\ref{DHFB}) replaces the exact one, i.e.
\begin{equation}
    \langle\hat{\cal O}\rangle={\rm Tr}({\cal D}_{\rm HFB}\hat{\cal O})~.
    \label{averageHFB}
    \end{equation}
The generalized particle-density matrix $R$ is related to the 
generalized quasiparticle-density matrix $Q$ through the Bogolyubov 
transformation 
(\ref{UV}) as
\begin{equation}
R={\cal U}^{\dagger}Q{\cal U}~,
\label{RUQU}
\end{equation}
where
\begin{equation}
    R=\left(\begin{array}{cc}
	{\rho}&{\tau}\\-{\tau}^{*}&{1}-\rho^{*}
	\end{array}\right)~,\hspace{5mm} 
    Q=\left(\begin{array}{cc}
	{q}&{t}\\-{t}^{*}&{1}-{q}^{*}
	\end{array}\right)
	=\left(\begin{array}{cc}
	n&{0}\\{0}&{1}-n\end{array}\right)~,
	\label{RQ}
	\end{equation}
with
\begin{equation}
	{\cal U}=\left(\begin{array}{cc}
	U^{*}&V^{*}\\V&U
	\end{array}\right)~,\hspace{5mm} {\cal U}{\cal 
	U}^{\dagger}={\bf 1}~.
	\label{U}
	\end{equation}
The matrix elements of the single-particle matrix $\rho$ and particle 
pairing 
tensor $\tau$ within the FT-HFB approximation are evaluated as
\begin{equation}
    \rho_{ij}=\langle a_{j}^{\dagger}a_{i}\rangle~,\hspace{5mm} 
    \tau_{ij}=\langle a_{j}a_{i}\rangle~,
    \label{rot}
    \end{equation}
while those of the quasiparticle matrix $q$ are given in terms 
of the quasiparticle occupation number since
\begin{equation}    
q_{ij}=\langle\alpha_{j}^{\dagger}\alpha_{i}\rangle=\delta_{ij}n_{i}~,
    \hspace{5mm} t_{ij}=\langle\alpha_{j}\alpha_{i}\rangle=0~,
\label{q}
\end{equation}
which follow from the HFB approximation (\ref{HHFB}).
Using the inverse transformation 
of 
(\ref{UV}), the particle densities are obtained 
as~\cite{Goodman1}
\begin{equation}
\rho={U}^{\rm T}nU^{*}+V^{\dagger}(1-n)V~,\hspace{6mm} 
\tau={U}^{\rm T}nV^{*}+V^{\dagger}(1-n)U~.
\label{rhot}
\end{equation}
By minimizing the grand potential $\Omega$ according to Eq. 
(\ref{minimum}), the FT-HFB equations were derived in 
the following form~\cite{Goodman1}
\begin{equation}
 \left(\begin{array}{cc}
	{\cal H}&\Delta\\-\Delta^{*}&-{\cal H}^{*}
	\end{array}\right) \left(\begin{array}{c}
	U_{i}\\ 
	V_{i}\end{array}\right)=E_{i} \left(\begin{array}{c}
	U_{i}\\ 
	V_{i}\end{array}\right)~,
	\label{FTHFBeq}
	\end{equation}
where
\begin{equation}
    {\cal H}={\cal T}-\mu+\Gamma~,\hspace{6mm} 
    \Gamma_{ij}=\sum_{kl}v_{ikjl}\rho_{lk}~,\hspace{6mm}
    \Delta_{ij}=\frac{1}{2}\sum_{kl}v_{ijkl}\tau_{kl}~.
    \label{HGammaDelta}
    \end{equation}
The total energy ${\cal E}$, entropy $S$, and particle number $N$ 
from Eq. (\ref{ESN})
are now given within the FT-HFB theory as
\begin{equation}
    {\cal E}={\rm Tr}[({\cal T}+\frac{1}{2}\Gamma)\rho+\frac{1}{2}\Delta 
\tau^{\dagger}]~,
    \label{Ecal}
\end{equation}
\begin{equation}
    S=-\sum_{i}[n_{i}{\rm ln}n_{i}+(1-n_{i}){\rm ln}(1-n_{i})]~,
    \label{Sentro}
\end{equation}
\begin{equation}
     N={\rm Tr}\rho~,
     \label{N}
\end{equation}
from which one can easily calculate the grand potential $\Omega$ 
(\ref{Omega}).

At zero temperature ($T=$ 0) the quasiparticle occupation number 
vanishes: $n_{i}=$0, and the average (\ref{averageHFB}) reduces to the average in the quasiparticle 
vacuum (\ref{qpvacuum}). The quasiparticle-density 
matrix $Q$ (\ref{RQ}) becomes
\begin{equation}
    Q(T=0)\equiv Q_{0}=\left(\begin{array}{cc}
	0&0\\0&1
	\end{array}\right)~,\hspace{6mm} {\rm for}\hspace{2mm} {\rm which}\hspace{5mm} Q_{0}^{2}=Q_{0}~.
	\label{Q0}
	\end{equation}
Therefore, for the generalized particle-density matrix 
$R_{0}=R(T=0)$ the following unitarity relation
holds 
\begin{equation}
    R_{0}^{2}=R_{0}~,\hspace{5mm} {\rm where}\hspace{5mm} R_{0}={\cal U}^{\dagger}Q_{0}{\cal U}~.
    \label{unitarity}
    \end{equation}
However, the idempotent (\ref{unitarity}) no longer holds at $T\neq$ 0. 
Indeed, from Eqs. (\ref{RUQU}) and (\ref{RQ}) it follows that 
\begin{equation}
    R-R^{2}={\cal U}^{\dagger}(Q-Q^{2}){\cal U}~,
    \label{R-R2}
    \end{equation}
    which leads to
    \begin{equation}
	{\rm Tr}(R-R^{2})={\rm 
	Tr}(Q-Q^{2})=2\sum_{i}n_{i}(1-n_{i})\equiv 2(\delta{\cal N})^{2}\neq 
	0~,\hspace{5mm} (T\neq 0)~.
	\label{TrR-R2}
	\end{equation}
The quantity $\delta{\cal N}^{2}=\sum_{i}n_{i}(1-n_{i})$ in 
Eq. (\ref{TrR-R2}) is nothing but the 
quasiparticle-number fluctuation since
\[
    \delta{\cal N}^{2}=\langle\hat{\cal N}^{2}
    \rangle-\langle\hat{\cal N}\rangle^{2}=\langle\sum_{i}\hat{\cal 
N}_{i}
    +\sum_{i\neq j}\hat{\cal N}_{i}
    \hat{\cal N}_{j}\rangle - \sum_{i}n_{i}^{2}-\sum_{i\neq 
    j}n_{i}n_{j}
    \]
  \begin{equation}  
    =\sum_{i}n_{i}(1-n_{i})  =\sum_{i}\delta{\cal N}_{i}^{2}~,
    \label{dN2}
    \end{equation}
    where $\delta{\cal N}_{i}^{2}=n_{i}(1-n_{i})$
	is the fluctuation of quasiparticle number on the $i$-th orbital.
Therefore, in order to restore the idempotent 
of type (\ref{unitarity}) at $T\neq$ 0 a new approximation should be found such that it includes 
the quasiparticle-number fluctuation in the 
quasiparticle-density matrix. 
\subsection{MHFB theory} 
Let us consider, instead of the FT-HFB density operator ${\cal D}_{\rm 
HFB}$ (\ref{DHFB}), an improved approximation, $\bar{\cal D}$, to the density 
operator ${\cal D}$. This approximated density operator $\bar{\cal D}$ 
should satisfy two following requirements:

(i) The average 
\begin{equation}
\langle\langle\hat{\cal O}\rangle\rangle\equiv{\rm Tr}(\bar{\cal D}\hat{\cal O}),
\label{averagebar}
\end{equation}
in which $\bar{\cal D}$ is used in place of ${\cal D}$ (or ${\cal D}_{\rm 
HFB}$), yields 
\begin{equation}
    \bar{R}={\cal U}^{\dagger}\bar{Q}{\cal U}~
    \label{RBARUQBARU}
    \end{equation}
for the 
Bogolyubov transformation ${\cal U}$ (\ref{U}), where one has the 
modified matrices
\begin{equation}
    \bar{R}=\left(\begin{array}{cc}
	\bar{\rho}&\bar{\tau}\\-\bar{\tau}^{*}&{1}-\bar\rho^{*}
	\end{array}\right)~,\hspace{5mm} 
    \bar{Q}=\left(\begin{array}{cc}
	\bar{q}&{t}\\-\bar{t}^{*}&{1}-\bar{q}^{*}
	\end{array}\right)~,
	\label{RQbar}
	\end{equation}
with	
\begin{equation}
    \bar{\rho}_{ij}=\langle\langle a_{j}^{\dagger}a_{i}\rangle\rangle~,\hspace{5mm} 
    \bar{\tau}_{ij}=\langle\langle a_{j}a_{i}\rangle\rangle~,
    \label{rotbar}
    \end{equation}
\begin{equation}    
\bar{q}_{ij}=\langle\langle\alpha_{j}^{\dagger}\alpha_{i}\rangle\rangle=
\delta_{ij}\bar{n}_{i}~,
    \hspace{5mm} 
 \bar{t}_{ij}=\langle\langle\alpha_{j}\alpha_{i}\rangle\rangle={\bf\Lambda}_{ij}~
\label{qbar}
\end{equation}
instead of matrices $R$ and $Q$ in Eqs. (\ref{RQ}), (\ref{rot}), and 
(\ref{q}).
The non-zero values of $\bar{t}_{ij}$ in 
Eq. (\ref{qbar}) are caused by the quasiparticle correlations in the 
thermal equilibrium, which are now included in the average 
$\langle\langle\ldots\rangle\rangle$ using the density operator 
$\bar{\cal D}$.

(ii) The modified quasiparticle-density matrix $\bar{Q}$ satisfies the 
unitarity relation   
\begin{equation}
(\bar{Q})^{2}=\bar{Q}~.
\label{Qbar2}
\end{equation}
The solution of Eq. (\ref{Qbar2}) immediately yields 
the matrix $\bf\Lambda$ in the canonical form
\begin{equation}
{\bf\Lambda}=\sqrt{\bar{n}(1-\bar{n})}\equiv
\left(\begin{array}{cccccccccccc}
	&&&&&&&&&&0&-\Lambda_{1}\\
	&&&&&&&&&&\Lambda_{1}&0\\
	&&&&&&&&0&-\Lambda_{2}&&\\
	&&&&&&&&\Lambda_{2}&0&&\\
	&&&&&&.&&&&\\
	&&&.&&&&&&&\end{array}\right)~,\hspace{5mm} 
	\Lambda_{i}=\sqrt{\bar{n}_{i}(1-\bar{n}_{i})}~.
\label{x}
\end{equation}
Comparing this result with Eq. (\ref{dN2}), it is clear 
that tensor ${\bf\Lambda}$ consists of the quasiparticle-number 
fluctuation $\delta\bar{\cal N}_{i}=\sqrt{\bar{n}_{i}(1-\bar{n}_{i})}$.
From Eq. (\ref{RBARUQBARU}) 
it is easy to see that the unitarity relation holds for the 
modified generalized single-particle density matrix $\bar{R}$ since
$
    \bar{R}-\bar{R}^{2}={\cal U}^{\dagger}(\bar{Q}-\bar{Q}^{2}){\cal U}=0~
$
due to Eq. (\ref{Qbar2}) and the unitary matrix ${\cal U}$. 

Let us define the modified-quasiparticle operators 
$\bar{\alpha}^{\dagger}_{i}$ and $\bar{\alpha}_{i}$, which behave in
the average (\ref{averagebar}) exactly as the usual quasiparticle 
operators 
$\alpha_{i}^{\dagger}$ and $\alpha_{i}$ do in the quasiparticle 
ground state, namely
\begin{equation}            
    \langle\langle\bar{\alpha}^{\dagger}_{i}\bar{\alpha}_{k}\rangle\rangle=
    \langle\langle\bar{\alpha}^{\dagger}_{i}\bar{\alpha}_{k}^{\dagger}\rangle\rangle=
    \langle\langle\bar{\alpha}_{k}\bar{\alpha}_{i}\rangle\rangle=0~.
    \label{avealphabar}
    \end{equation}
In the same way as for the usual Bogolyubov transformation (\ref{UV}),
we search for a transformation between these modified-quasiparticle 
operators ($\bar{\alpha}^{\dagger}_{i}$, $\bar{\alpha}_{i}$) and the 
usual quasiparticle ones  (${\alpha}^{\dagger}_{i}$, ${\alpha}_{i}$)
in the following form 
\begin{equation}
	\left( \begin{array}{c}\bar{\alpha}^{\dagger}\\
	\bar{\alpha}\end{array}
	\right)=
	\left(\begin{array}{cc}
	w&z\\z^{*}&w^{*}
	\end{array}\right)
	\left(\begin{array}{c}
	\alpha^{\dagger}\\ 
	\alpha\end{array}\right)~,
     \label{Bogo2}
     \end{equation}
with the unitary property similar to Eq. (\ref{UVproperties}) 
for $U$ and $V$ matrices $:
    ww^{\dagger}+zz^{\dagger}={\bf 1}~.
$
Using the inverse transformation of (\ref{Bogo2}) and the 
requirement (\ref{avealphabar}), we obtain
\begin{equation}
    \bar{n}_{i}=\langle\langle\alpha^{\dagger}_{i}\alpha_{i}\rangle\rangle= 
    \sum_{k}z_{ik}z_{ik}^{*}~.
    \label{nz}
    \end{equation}
From this equation and the unitarity condition (\ref{unitarity}), it 
follows that $zz^{\dagger}=\bar{n}$ and $ww^{\dagger}={\bf 1}-\bar{n}$. Since 
${\bf 1}-\bar{n}$ and $\bar{n}$ are real diagonal matrices, 
the canonical form of matrices $w$ and $z$ is found as
\begin{equation}    
    \label{wzmatrices}  
    w=\left(\begin{array}{ccccccccccc}
	w_{1}&0&&&&&&&&&\\
	0&w_{1}&&&&&&&&&\\
	&&w_{2}&0&&&&&&&\\
	&&0&w_{2}&&&&&&&\\
	&&&&.&&&&&\\
	&&&&&&&.&&\end{array}\right)~,\hspace{4mm} 
	z=\left(\begin{array}{ccccccccccc}
	&&&&&&&&&0&-z_{1}\\
	&&&&&&&&&z_{1}&0\\
	&&&&&&&0&-z_{2}&&\\
	&&&&&&&z_{2}&0&&\\
	&&&&&&.&&&&\\
	&&&&.&&&&&&\end{array}\right)~, 
    \end{equation}
    where
	$w_{i}=\sqrt{1-\bar{n}_{i}}$, $z_{i}=\sqrt{\bar{n}_{i}}$.

We now show that we can obtain the idempotent 
$\bar{R}^{2}=\bar{R}$ by applying 
the secondary Bogolyubov 
transformation (\ref{Bogo2}), which automatically leads to Eq. 
(\ref{Qbar2}).  
Indeed, using the inverse transformation of (\ref{Bogo2}) with matrices 
$w$ and $z$ given in 
Eq. (\ref{wzmatrices}), we found that the 
modified quasiparticle-density matrix $\bar{Q}$ can be obtained as
\begin{equation}
    {\cal W}^{\dagger}\bar{Q}_{0}{\cal W}=\left(\begin{array}{cc}
	\bar{n}&[\sqrt{\bar{n}(1-\bar{n})}]
	^{\dagger}\\\sqrt{\bar{n}(1-\bar{n})}&1-\bar{n}\end{array}\right)
	    \equiv\bar{Q}~,
	\label{WQ0W}
\end{equation}
where
\begin{equation}
    {\cal W}=\left(\begin{array}{cc}	
(\sqrt{1-\bar{n}})^{*}&(\sqrt{\bar{n}})^{*}\\\sqrt{\bar{n}}&\sqrt{1-\bar{n}}
\end{array}\right)~,	
\hspace{5mm} {\cal W}{\cal W}^{\dagger}={\bf 1}~,
\label{W}
\end{equation}
and
\begin{equation}
    \bar{Q}_{0}=\left(\begin{array}{cc}	
\langle\langle\bar{\alpha}^{\dagger}\bar{\alpha}\rangle\rangle
&\langle\langle\bar{\alpha}\bar{\alpha}\rangle\rangle\\
\langle\langle\bar{\alpha}^{\dagger}\bar{\alpha}^{\dagger}
\rangle\rangle&1-\langle\langle\bar{\alpha}^{\dagger}\bar{\alpha}\rangle\rangle
\end{array}\right)=
\left(\begin{array}{cc}	
0&0\\0&1
\end{array}\right)~,\hspace{5mm} \bar{Q}_{0}^{2}=\bar{Q}_{0}~,
\label{Q0bar}
\end{equation}
due to Eq. (\ref{avealphabar}). This result shows another way
of deriving the
modified quasiparticle-density matrix $\bar{Q}$ (\ref{RQbar})
from the density matrix $\bar{Q}_{0}$ of the modified quasiparticles
($\bar{\alpha}^{\dagger}_{i}$, $\bar{\alpha}_{i}$). This matrix 
$\bar{Q}_{0}$ is identical to
the zero-temperature quasiparticle-density matrix $Q_{0}$ (\ref{Q0}).
Substituting this result into the right-hand side (rhs) of 
Eq. (\ref{RBARUQBARU}), we obtain
\begin{equation}
\bar{R}=\bar{\cal U}^{\dagger}\bar{Q}_{0}\bar{\cal U}~,
\label{RUBARQ0UBAR}
\end{equation}
where
\[
\bar{\cal U}={\cal W}{\cal U}=
 \left(\begin{array}{cc}	
\bar{U}^{*}&\bar{V}^{*}\\\bar{V}&\bar{U}\end{array}\right)=
\]
\begin{equation}
 \left(\begin{array}{cc}	
(\sqrt{1-\bar{n}})^{*}U^{*}+(\sqrt{\bar{n}})^{*}V
&~~~~(\sqrt{1-\bar{n}})^{*}V^{*}+(\sqrt{\bar{n}})^{*}U\\
\sqrt{1-\bar{n}}V+\sqrt{\bar{n}}U^{*}&~~~~\sqrt{1-\bar{n}}U+\sqrt{\bar{n}}V^{*}\end{array}\right)~.
\label{Ubar}
 \end{equation}
This equation is the generalized form of the modified Bogolyubov 
coefficients $\bar{u}_{j}$ and $\bar{v}_{j}$ given in 
Eq. (38) of Ref. \cite{MBCS}.  From Eqs. 
(\ref{U}), (\ref{W}), and (\ref{Ubar}), it follows that 
$\bar{\cal U}\bar{\cal U}^{\dagger}={\bf 1}$, i.e.  
transformation (\ref{RUBARQ0UBAR}) is unitary.
Therefore, from the idempotent (\ref{Q0bar}) it follows 
that $\bar{R}^{2}=\bar{R}$. 

Applying the Wick's theorem for the ensemble 
average, one obtains 
the expressions for the modified 
total energy $\bar{\cal E}$
\begin{equation}
\bar{\cal E}={\rm 
Tr}[({\cal T}+\frac{1}{2}\bar{\Gamma})\bar{\rho}+\frac{1}{2}
\bar{\Delta}\bar{\tau}^{\dagger}]~,
\label{Ecalbar}
\end{equation}
where
\begin{equation}
\bar{\Gamma}_{ij}=\sum_{kl}v_{ikjl}\bar{\rho}_{lk}~,\hspace{2mm} 
\bar{\Delta}_{ij}=\frac{1}{2}\sum_{kl}v_{ijkl}\bar{\tau}_{kl}~.
\label{Deltabar}
\end{equation}
From Eq. (\ref{RUBARQ0UBAR}) we obtain the modified single-particle 
density matrix $\bar{\rho}$ and modified
particle-pairing tensor $\bar{\tau}$ in the following form
\begin{equation}
\bar{\rho}={U}^{\rm T}\bar{n}U^{*}+V^{\dagger}(1-\bar{n})V+
{U}^{\rm 
T}\bigg[\sqrt{\bar{n}(1-\bar{n})}\bigg]^{\dagger}V+V^{\dagger}\sqrt{\bar{n}(1-
\bar{n})}U^{*}~,
\label{rhobar}
\end{equation}
\begin{equation}
\bar{\tau}={U}^{\rm T}\bar{n}V^{*}+V^{\dagger}(1-\bar{n})U
+{U}^{\rm 
T}\bigg[\sqrt{\bar{n}(1-\bar{n})}\bigg]^{\dagger}U+V^{\dagger}\sqrt{\bar{n}(1-\bar{n})}
V^{*}~.
\label{taubar}
\end{equation}
As compared to
Eq. (\ref{rhot}) within the FT-HFB 
approximation, Eqs. (\ref{rhobar}) and (\ref{taubar}) 
contain the last two terms  
$\sim[\sqrt{\bar{n}(1-\bar{n})}]^{\dagger}$ and $\sim\sqrt{\bar{n}(1-\bar{n})}$, 
which arise due to quasiparticle-number 
fluctuation. Also the quasiparticle occupation number is now 
$\bar{n}$ [See Eq. (\ref{qbar})]
instead of $n$ (\ref{ni}).

We derive the MHFB equations following the same 
variational procedure, which was used to derive the FT-HFB 
equations in Ref. \cite{Goodman1}. According it, we minimize 
the grand potential $\delta{\bar{\Omega}}=$ 0 by varying $U$, $V$, and 
$\bar{n}$, where 
\begin{equation}
\bar{\Omega}=\bar{\cal 
E}-T\bar{S}-\bar{\mu}N~.
\label{Omegabar}
\end{equation}
The MHFB 
equations formally look like the FT-HFB ones, namely 
(\ref{FTHFBeq})
\begin{equation}
 \left(\begin{array}{cc}
	\bar{\cal H}&\bar{\Delta}\\-\bar{\Delta}^{*}&-\bar{\cal H}^{*}
	\end{array}\right) \left(\begin{array}{c}
	U_{i}\\ 
	V_{i}\end{array}\right)=\bar{E}_{i} \left(\begin{array}{c}
	U_{i}\\ 
	V_{i}\end{array}\right)~,
	\label{MHFBeq}
	\end{equation}
where, however
\begin{equation}
    \bar{\cal H}={\cal T}-\bar{\mu}+\bar{\Gamma}~
    \label{Hbar}
    \end{equation}
with $\bar{\Gamma}$ and $\bar{\Delta}$ given by Eq. (\ref{Deltabar}).
 The equation for particle number $N$ within the MHFB theory is
 \begin{equation}
     N={\rm Tr}\bar{\rho}~.
     \label{NMHFB}
     \end{equation}
By solving Eq. (\ref{MHFBeq}), one obtains the modified quasiparticle energy 
$\bar{E}_{i}$, which is different from $E_{i}$ 
in Eqs (\ref{FTHFBeq}) due to 
the change of the HF and pairing potentials. Hence, 
the MHFB quasiparticle Hamiltonian $H_{\rm MHFB}$ can be 
written as
\begin{equation}
    H-\bar{\mu}\hat{N}\approx H_{\rm MHFB}=\bar{E}_{0}+\sum_{i}\bar{E}_{i}\hat{\cal N}_{i}~,
    \label{HMHFB}
    \end{equation}
    instead of (\ref{HHFB}).
This implies that
the approximated density operator 
$\bar{D}$ (\ref{averagebar}) within the MHFB theory 
can be represented in the form similar 
to (\ref{DHFB}), namely
\begin{equation}
 \bar{\cal D}\equiv{\cal D}_{\rm MHFB}=
 \prod_{i}[\bar{n}_{i}\hat{\cal N}_{i}+(1-\bar{n}_{i})
 (1-\hat{\cal N}_{i})].~
 \label{DMHFB}
 \end{equation}  
From here it follows that the formal expression for the modified entropy 
$\bar{S}$ is the same as that given in Eq. (\ref{Sentro}), i.e.
\begin{equation}
    \bar{S}=-\sum_{i}[\bar{n}_{i}{\rm ln}\bar{n}_{i}+(1-\bar{n}_{i}){\rm 
    ln}(1-\bar{n}_{i})]~,
    \label{barS}
\end{equation}
Using the thermodynamic definition of temperature in terms of entropy 
$1/T=\delta{\bar{S}}/\delta{\bar{\cal E}}$ and carrying out the 
variation over $\delta\bar{n}_{i}$, we find
\begin{equation}
\frac{\delta\bar{\cal E}}{\delta\bar{n}_{i}}\equiv
\bar{E}_{i}=T\frac{\delta\bar{S}}{\delta\bar{n}_{i}}=
T{\rm ln}\bigg(\frac{1-\bar{n}_{i}}{\bar{n}_{i}}\bigg)~.
\label{varE}
\end{equation}
Inverting Eq. (\ref{varE}), we obtain
\begin{equation}
    \bar{n}_{i}=\frac{1}{{\rm e}^{\beta\bar{E}_{i}}+1}~.
    \label{nbar}
    \end{equation}
This result shows that the functional dependence of 
quasiparticle occupation number 
$\bar{n}_{i}$ on quasiparticle energy and temperature within the MHFB theory 
is also given by the 
Fermi-Dirac distribution of noninteracting quasiparticles but with 
the modified energies $\bar{E}_{i}$ defined by the MHFB equations 
(\ref{MHFBeq}). 
Therefore we will omit the bar over 
$\bar{n}_{i}$ and use 
the same Eq. (\ref{ni}) with ${E}_{i}$ replaced
with $\bar{E}_{i}$ for the MHFB equations.  
 \subsection{MBCS theory}
 In the limit with equal pairing matrix elements 
 $G_{ij}=G$, 
 neglecting the contribution of
 $G$ to the HF potential so that $\bar{\Gamma}=$ 0, the HF Hamiltonian
 becomes
 \begin{equation}
     \bar{\cal H}_{ij}=(\epsilon_{i}-\bar{\mu})\delta_{ij}~.
     \label{HHFBCS}
     \end{equation}
 The pairing potential (\ref{Deltabar}) takes now the simple form 
 \begin{equation}
      \bar{\Delta}=-G\sum_{k>0}\bar{\tau}_{k\widetilde{k}}~.
     \label{DeltaMBCS}
     \end{equation}
 The Bogolyubov transformation (\ref{UV}) for spherical nuclei 
reduces to
 \[
 {\alpha}_{jm}^{\dagger}=u_{j}a_{jm}^{\dagger}+v_{j}(-)^{j+m}a_{j-m}~,
 \]\begin{equation}
 (-)^{j+m}{\alpha}_{j-m}=u_{j}(-)^{j+m}
 a_{j-m}-v_{j}a_{jm}^{\dagger}~,
 \label{uv}
 \end{equation}
 while the secondary Bogolyubov transformation (\ref{Bogo2}) 
 becomes~\cite{MBCS}
 \[
\bar{\alpha}_{jm}^{\dagger}=\sqrt{1-n_{j}}\alpha_{jm}^{\dagger}-\sqrt{n_{j}}
(-)^{j+m}{\alpha}_{j-m}~,
 \]
 \begin{equation}
 (-)^{j+m}\bar{\alpha}_{j-m}=\sqrt{1-n_{j}}(-)^{j+m}
 {\alpha}_{j-m}+\sqrt{n_{j}}{\alpha}_{jm}^{\dagger}~.
 \label{w}
 \end{equation}
 The $U$, $V$, $1-n$, $n$, and $\sqrt{n(1-n)}$ matrices are 
 now block diagonal in each two-dimensional subspace spanned by the 
 quasiparticle state $|j\rangle$ 
 and its time-reversal partner 
$|\widetilde{j}\rangle=(-)^{j+m}|j-m\rangle$
 \begin{equation}
  U= \left(\begin{array}{cc}
	u_{j}&0\\0&u_{j}
	\end{array}\right)~,\hspace{5mm}  V= \left(\begin{array}{cc}
	0&v_{j}\\-v_{j}&0
	\end{array}\right)~,
	\label{UiVi}
	\end{equation}
\[
  1-n= \left(\begin{array}{cc}
	1-n_{j}&0\\0&1-n_{j}
	\end{array}\right)~,\hspace{5mm}  n= \left(\begin{array}{cc}
	n_{j}&0\\0&n_{j}
	\end{array}\right)~,\]
\begin{equation} \sqrt{n(1-n)}= 
	\left(\begin{array}{cc}
	0&-\sqrt{n_{j}(1-n_{j})}\\\sqrt{n_{j}(1-n_{j})}&0
	\end{array}\right)~,
	\label{nmatrices}
	\end{equation}
Substituting these matrices into the rhs of Eqs. (\ref{rhobar}) 
and (\ref{taubar}), 
we find
\begin{equation}
\bar{\rho}_{j~\widetilde{j}}=(1-2n_{j})v_{j}^{2}+n_{j}-2\sqrt{n_{j}(1-n_{j})}
u_{j}v_{j}~,
\label{rhoMBCS}
\end{equation}
\begin{equation}
\bar{\tau}_{j~\widetilde{j}}=-(1-2n_{j})u_{j}v_{j}+\sqrt{n_{j}(1-n_{j})}(u_{j}^{2}-
v_{j}^{2})~.
\label{tauMBCS}
\end{equation}
Substituting now Eqs. (\ref{tauMBCS}) and (\ref{rhoMBCS}) into the 
rhs of Eqs. 
(\ref{DeltaMBCS}) and (\ref{NMHFB}), respectively, we obtain the MBCS 
equations for spherical nuclei in the following form:
\begin{equation}
 \bar{\Delta}=G\sum_{j}\Omega_{j} 
 [(1-2n_{j})u_{j}v_{j}-\sqrt{n_{j}(1-n_{j})}(u_{j}^{2}-v_{j}^{2})]~,
 \label{MBCSgap}
 \end{equation}
 \begin{equation}    
N=2\sum_{j}\Omega_{j}[(1-2n_{j})v_{j}^{2}+n_{j}-2\sqrt{n_{j}(1-n_{j})}
    u_{j}v_{j}]~.
 \label{MBCSN}
 \end{equation}
Comparing the conventional FT-BCS equations, we see that 
the MBCS equations explicitly include the effect of 
quasiparticle-number fluctuation $\sim\delta{\cal N}_{j}$ 
in the last terms at their rhs, which are the thermal gap 
$-G\sum_{j}\Omega_{j}\sqrt{n_{j}(1-n_{j})}(u_{j}^{2}-v_{j}^{2})$, 
and the thermal-fluctuation of particle number 
$\delta N=\sum_{j}\delta N_{j}=
-4\sum_{j}\Omega_{j}u_{j}v_{j}(\delta{\cal N}_{j})$ 
in Eq. (\ref{MBCSN}).
These terms are ignored within 
the FT-BCS theory. Hence Eqs. (\ref{MBCSgap}) and 
(\ref{MBCSN}) show for the first time how the effect of statistical 
fluctuations is included in the MBCS (MHFB) theory at finite 
temperature 
on a microscopic ground.
So far this effect was treated only within the framework
of the macroscopic Landau theory of phase transition~\cite{Moretto}. 
\section{Phonon-damping model in quasiparticle representation}
The quasiparticle representation of the PDM 
Hamiltonian~\cite{PDM1} 
is obtained by adding the superfluid pairing interaction and 
expressing the particle ($p$) and hole ($h$) creation and 
destruction operators, $a_{s}^{\dagger}$ and $a_{s}$ ($s=p, h$), in 
terms of the quasiparticle operators, $\alpha_{s}^{\dagger}$ and
$\alpha_{s}$, using the Bogolyubov's canonical transformation.
As a result, the PDM Hamiltonian 
for the description of E$\lambda$ excitations 
can be written in spherical basis as
\[
H=\sum_{jm}E_{j}\alpha_{jm}^{\dagger}\alpha_{jm}+
\sum_{\lambda\mu i}\omega_{\lambda i}b_{\lambda\mu 
i}^{\dagger}b_{\lambda\mu i}~+
\]
\begin{equation}
\frac{1}{2}\sum_{\lambda\mu i}\frac{(-)^{\lambda-\mu}}{\hat{\lambda}}
\sum_{jj'}f_{jj'}^{(\lambda)}\Biggl\{u_{jj'}^{(+)}
\biggl[A_{jj'}^{\dagger}(\lambda\mu)+A_{jj'}(\lambda\tilde{\mu})\biggr]+
\label{H}
\end{equation}
\[
v_{jj'}^{(-)}
\biggl[B_{jj'}^{\dagger}(\lambda\mu)+B_{jj'}(\lambda\tilde{\mu})\biggr]\Biggr\}
\biggl(b_{\lambda\mu i}^{\dagger}+b_{\lambda\tilde{\mu} i}\biggr)~,
\]
where $\hat{\lambda}=\sqrt{2\lambda+1}$. 
The first term at the rhs of Hamiltonian (\ref{H})
corresponds to the independent-quasiparticle field. The second term 
stands for the phonon field described by phonon operators, 
$b_{\lambda\mu i}^{\dagger}$ and $b_{\lambda\mu i}$, with 
multipolarity $\lambda$, which generate the harmonic collective
vibrations such as GDR. Phonons are ideal bosons within the PDM, i.e. 
they have no fermion structure. The last term is the coupling between 
quasiparticle and phonon fields, 
which is responsible for the microscopic damping
of collective excitations. 
 
In Eq. (\ref{H}) the following standard notations are used
\begin{equation} 
A_{jj'}^{\dagger}(\lambda\mu)=\sum_{mm'}\langle 
jmj'm'|\lambda\mu\rangle\alpha_{jm}^{\dagger}\alpha_{j'm'}^{\dagger}~,
\label{Adagger}
\end{equation}
\begin{equation} 
B_{jj'}^{\dagger}(\lambda\mu)=-\sum_{mm'}(-)^{j'-m'}\langle 
jmj'-m'|\lambda\mu\rangle\alpha_{jm}^{\dagger}\alpha_{j'm'}~,
\label{Bdagger}
\end{equation}
with
$(\lambda\tilde{\mu})\longleftrightarrow(-)^{\lambda-\mu}(\lambda-\mu)$.
Functions $u_{jj'}^{(+)}\equiv u_{j}v_{j'}+v_{j}u_{j'}$ and 
$v_{jj'}^{(-)}
\equiv u_{j}u_{j'}-v_{j}v_{j'}$ are combinations of Bogolyubov's $u$ 
and $v$ coefficients. The quasiparticle energy $E_{j}$ is 
calculated from the single-particle energy $\epsilon_{j}$ as
\begin{equation}
E_{j}=\sqrt{({\epsilon}_{j}'-\epsilon_{\rm F})^{2}+\Delta^{2}}~,
\hspace{1cm} {\epsilon}_{j}'\equiv \epsilon_{j}-Gv_{j}^{2},
\label{epsilonj}
\end{equation}
where the pairing gap $\Delta$ and the Fermi energy $\epsilon_{\rm F}$  
are defined as solutions of the BCS equations. At $T\neq$ 0
the thermal pairing gap $\Delta(T)$ (or $\bar\Delta(T)$) is defined from the
finite-temperature BCS (or MBCS) equations.

The equation for the propagation of the GDR phonon, which is damped 
due to coupling to the quasiparticle field, is derived making use of 
the double-time Green's function method (introduced by
Bogolyubov and Tyablikov, and developed further by Zubarev~\cite{Bogo}). Following the 
standard procedure of deriving the equation for the double-time 
retarded Green's function with respect to the Hamiltonian (\ref{H}), one obtains 
a closed set of equations for the Green's functions for phonon and
quasiparticle propagators. Making the 
Fourier transform into the energy plane $E$, and expressing all the 
Green functions in the set in terms of the one-phonon propagation 
Green function, we obtain the equation for the latter, 
$G_{\lambda i}(E)$, in the form
\begin{equation}
G_{\lambda i}(E)=\frac{1}{2\pi}\frac{1}{E-\omega_{\lambda 
i}-P_{\lambda i}(E)}~,
\label{GE}
\end{equation}
where the explicit form of 
the polarization operator $P_{\lambda i}(E)$
is
\[
P_{\lambda i}(E)=\frac{1}{\hat{\lambda}^{2}}\sum_{jj'}[f_{jj'}^{(\lambda)}]^{2}
\biggl[\frac{(u_{jj'}^{(+)})^{2}(1-n_{j}-n_{j'})(\epsilon_{j}+\epsilon_{j'})}
{E^{2}-(\epsilon_{j}+\epsilon_{j'})^{2}}~~
-\]
\begin{equation}
\frac{(v_{jj'}^{(-)})^{2}(n_{j}-n_{j'})(\epsilon_{j}-\epsilon_{j'})}
{E^{2}-(\epsilon_{j}-\epsilon_{j'})^{2}}\biggr]~.
\label{PE}
\end{equation}
The polarization operator (\ref{PE}) appears due to $ph$ -- 
phonon coupling in the last term of the rhs of Hamiltonian (\ref{H}).
The phonon damping $\gamma_{\lambda i}(\omega)$ ($\omega$ real) is 
obtained as the imaginary part of the analytic continuation of
the polarization operator $P_{\lambda i}(E)$ into the complex energy 
plane $E=\omega\pm i\varepsilon$. Its final form is
\[
\gamma_{\lambda i}(\omega)=\frac{\pi}{2\hat{\lambda}^{2}}
\sum_{jj'}[f_{jj'}^{(\lambda)}]^{2}\times
\]
\begin{equation}
\biggl\{
(u_{jj'}^{(+)})^{2}(1-n_{j}-n_{j'})
[\delta(E-E_{j}-E_{j'})-
\delta(E+E_{j}+E_{j'})]-
\label{gamma}
\end{equation}
\[
(v_{jj'}^{(-)})^{2}(n_{j}-n_{j'})[\delta(E-E_{j}+E_{j'})
-\delta(E+\epsilon_{j}-\epsilon_{j'})]\biggr\}.
\]
The energy $\bar{\omega}$ 
of giant resonance (damped collective phonon) is found 
as the solution of the equation: 
$\bar{\omega}-\omega_{\lambda i}-P_{\lambda i}(\bar{\omega})=0$~.
The width $\Gamma_{\lambda}$ of giant resonance 
is calculated as twice of the damping 
$\gamma_{\lambda}(\omega)$ at $\omega=\bar{\omega}$,
where $\lambda=$ 1 corresponds to the GDR width $\Gamma_{\rm GDR}$. 
The latter has the form
\[
\Gamma_{\rm GDR}=2\pi\biggl\{ F_{1}^{2}\sum_{ph}[u_{ph}^{(+)}]^{2}(1-n_{p}-n_{h})
\delta(E_{\rm GDR}-E_{p}-E_{h})~ + 
\]
\begin{equation}
F_{2}^{2}\sum_{s>s'}[v_{ss'}^{(-)}]^{2}(n_{s'}-n_{s})
\delta(E_{\rm GDR}-E_{s}+E_{s'})\biggr\}~,
\label{width}
\end{equation}
where $(ss')=(pp')$ and $(hh')$ with $p$ and $h$ denoting the orbital 
angular momenta $j_{p}$ and $j_{h}$ for particles and holes, 
respectively. The first sum at the rhs of Eq. (\ref{width}) is the 
quantal width $\Gamma_{\rm Q}$, which comes from 
the couplings of quasiparticle pairs 
$[\alpha_{p}^{\dagger}\otimes\alpha^{\dagger}_{h}]_{LM}$ 
to the GDR. At zero pairing they correspond to the couplings of 
$ph$ pairs, $[a^{\dagger}_{p}\otimes{{a}}_{\widetilde{h}}]_{LM}$ to the GDR.
The second sum comes from the coupling of  
$[\alpha_{s}^{\dagger}\otimes{{\alpha}}_{\widetilde{s'}}]_{LM}$ to the
GDR, and is called the thermal width $\Gamma_{\rm T}$ as it appears only at $T\neq$ 0.
At zero pairing they are 
$pp$ ($hh$) pairs, 
$[a^{\dagger}_{s}\otimes{a}_{\widetilde{s'}}]_{LM}$ (The tilde $~{\widetilde{}}$~ denotes 
the time-reversal operation). 

The line shape of the GDR is described by the strength function 
$S_{\rm GDR}(\omega)$, which is   
derived from the spectral intensity 
in the standard way using the analytic continuation of the
Green function (\ref{GE}) and
by expanding the polarization operator (\ref{PE}) 
around $\omega=E_{\rm GDR}$. The final form 
of $S_{\rm GDR}(\omega)$ is~\cite{PDM,PDM1}
\begin{equation}
S_{\rm GDR}(\omega)=\frac{1}{\pi}\frac{\gamma_{\rm GDR}(\omega)}
{(\omega-E_{\rm GDR})^{2}+\gamma_{\rm GDR}^{2}(\omega)}~.
\label{S}
\end{equation} 
The PDM is based on the following assumptions:

a1) The matrix elements for the coupling of GDR to non-collective $ph$ 
configurations, which causes the quantal width $\Gamma_{\rm Q}$, are all equal to $F_{1}$. 
Those for the coupling of GDR to $pp$ ($hh$), which causes the thermal 
width $\Gamma_{\rm T}$, 
are all equal to $F_{2}$.

a2) It is well established that 
the microscopic mechanism of the quantal (spreading) width 
$\Gamma_{\rm Q}$ comes from 
quantal coupling of $ph$ configurations to more complicated ones, 
such as $2p2h$ ones. The calculations performed 
in Refs. \cite{NFT} 
within two independent microscopic models, where such couplings to 
$2p2h$ configurations were explicitly included, have shown that $\Gamma_{\rm Q}$ 
depends weakly on $T$. Therefore, 
in order to avoid complicate numerical calculations, which are not 
essential for the increase of $\Gamma_{\rm GDR}$ 
at $T\neq$ 0, such microscopic mechanism is not 
included within PDM, assuming
that $\Gamma_{\rm Q}$ at $T=$ 0 is known. The model parameters
are then chosen so that the calculated $\Gamma_{\rm Q}$ and $E_{\rm 
GDR}$ reproduce the corresponding experimental values 
at $T=$ 0. 

Within assumptions (a1) and (a2) the model has only three 
$T$-independent parameters, which are the unperturbed phonon energy 
$\omega_{q}$, $F_{1}$, and  $F_{2}$. 
The parameters $\omega_{q}$ and $F_{1}$ are 
chosen so that after the $ph$-GDR coupling is switched on,
the calculated GDR energy $E_{\rm GDR}$ and width $\Gamma_{\rm GDR}$ 
reproduce the corresponding experimental values for GDR on 
the ground-state. At $T\neq$ 0, the coupling to $pp$ and $hh$ 
configurations is activated.
The $F_{2}$ parameter is then fixed at $T=$ 0 
so that the GDR energy $E_{\rm GDR}$ 
does not change appreciably with $T$. 
\section{Numerical results}

\subsection{Temperature dependence of pairing gap}
\begin{figure}                                                             
    \hspace{0.8cm} \includegraphics[width=10cm]{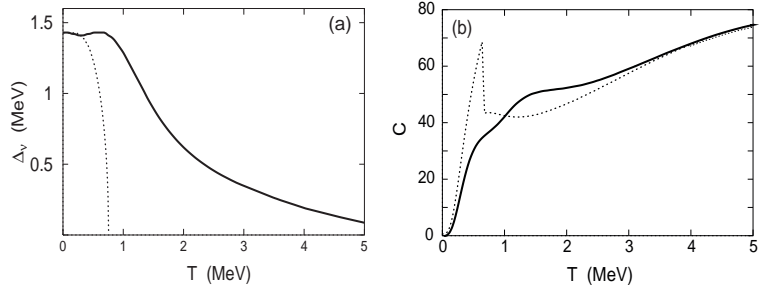}
\caption{\label{gapSn}Neutron pairing gap (a) and heat capacity (b)
for $^{120}$Sn as functions of $T$.
Solid and dotted lines show the MBCS and BCS 
gaps, respectively.}
\end{figure}
Shown in Fig. \ref{gapSn} (a) is the temperature dependence of the neutron 
pairing gap $\bar{\Delta}_{\nu}$ for $^{120}$Sn, which is 
obtained from the MBCS equation (\ref{MBCSgap}) using the 
single-particle energies 
determined within the Woods-Saxon potential at $T=$ 0. 
The pairing parameter $G_{\nu}$ is chosen to be equal to 0.13 MeV, which 
yields $\bar{\Delta}(T=0)\equiv\bar{\Delta}(0)\simeq$ 1.4 MeV. 
Contrary to the BCS gap (dotted line), 
which collapses at $T_{\rm c}\simeq$ 
0.79 MeV, the 
gap $\bar{\Delta}$ (solid line) does not vanish, but
decreases monotonously with increasing $T$ at $T\geq$ 1 MeV 
resulting in a long tail up to $T\simeq$ 5 MeV.
This behavior is caused by the thermal fluctuation 
of quasiparticle number in the MBCS equations (\ref{MBCSgap}). As the 
result, the heat capacity [Fig. \ref{gapSn} (b)] has no divergence at 
$T_{c}$, which is seen within the BCS theory.
\subsection{Temperature dependence of GDR width}
\begin{figure}                                                             
   \hspace{0.8cm} \includegraphics[width=10cm]{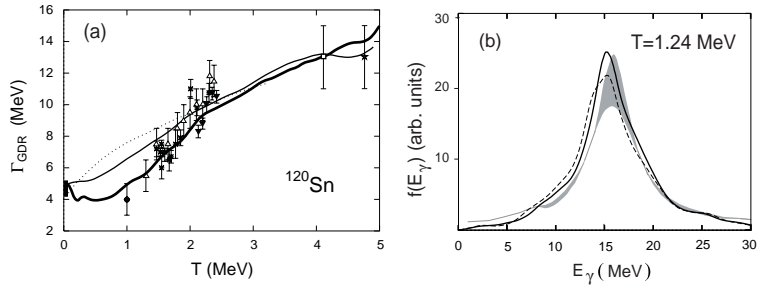}
\caption{\label{widthSn}(b): GDR width $\Gamma_{\rm GDR}$ as a function of 
$T$ for $^{120}$Sn. The thin and thick solid lines show the 
PDM results obtained neglecting pairing and  
including the renormalized gap 
$\widetilde{\Delta}=[1+1/\delta N^2]\bar{\Delta}$, respectively. 
The gap $\widetilde\Delta$ includes the correction 
$\delta N^{2}=\bar{\Delta}(0)^{2}\sum_{j}(j+1/2)/
[(\epsilon_{j}-\bar{\epsilon}_{\rm F})^{2}+\bar{\Delta}(0)^{2}]$ due to an
approximate number projection.
The prediction by the TFM is shown as the 
dotted line~$^{16}$; (b): GDR strength function at $T=$ 1.24 MeV. The 
dashed and solid lines show the results obtained without and including
the gap $\widetilde{\Delta}$, while experimental results are shown as the shaded area.}
\end{figure}

The GDR widths as a function of $T$ for $^{120}$Sn obtained within the
PDM are compared in Fig. \ref{widthSn} (a) with the experimental data and the prediction by the
thermal fluctuation model (TFM)~\cite{TFM}. 

The TFM interprets the
broadening of the GDR width via an adiabatic coupling of GDR 
to quadrupole deformations induced by thermal 
fluctuations. Even when thermal pairing is
neglected the PDM prediction, (the thin solid line) is already better than 
that given by the TFM, including the region of high $T$ where the
width's saturation is reported. The increase of the total width with
$T$ is driven by the increase of the thermal width $\Gamma_{\rm T}$, 
which is caused by coupling to $pp$ and $hh$
configurations, since the quantal width $\Gamma_{\rm Q}$ 
is found to decrease slightly with increasing $T$~\cite{PDM}. 
The inclusion of thermal pairing,
which yields a sharper Fermi surface, compensates
the smoothing of the Fermi surface with increasing $T$. This leads to 
a much weaker $T$-dependence of the GDR width at low $T$. 
As a result, the values of the width predicted by the
PDM in this region significantly drop (the thick solid line),
recovering the data point at $T=$ 1 MeV. The GDR strength function
obtained including the MBCS gap is also closer to the experimental
data than that obtained neglecting the thermal gap [Fig. \ref{widthSn}
(b)]. 

\begin{figure}                                                             
    \hspace{0.8cm} \includegraphics[width=9.5cm]{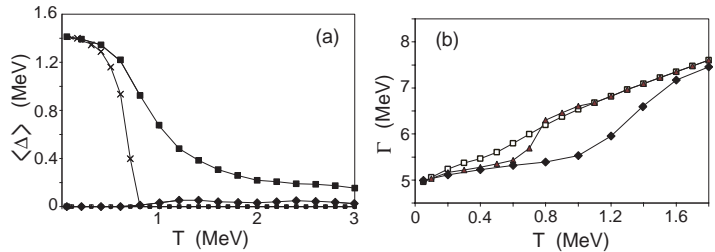}
\caption{(a): Pairing gaps for $^{120}$Sn averaged over thermal shape 
fluctuations versus $T$. 
Lines with triangles and crosses are the usual BCS proton and neutron 
pairing gaps, respectively, while those
with diamonds and squares denote the corresponding pairing gaps, which 
also include thermal fluctuations of pairing fields.
(b): GDR widths for $^{120}$Sn versus $T$. Open squares, 
triangles, and diamonds denote the widths obtained without pairing, 
including BCS pairing, and thermally fluctuating pairing field from (a), 
respectively.\label{Sn120}}
\end{figure}
The results discussed above have 
also been confirmed by our recent calculations within a macroscopic approach, which takes 
pairing fluctuations into account along with the thermal shape 
fluctuations~\cite{Aru}. Here the free energies 
are calculated using the Nilsson-Strutinsky method at $T\neq$
0, including thermal pairing correlations.  
The GDR is coupled  to the nuclear shapes through a simple 
anisotropic harmonic oscillator model with a separable dipole-dipole 
interaction.  The observables are averaged over the shape parameters and pairing gap. 
Our study reveals that the observed quenching of GDR width at low $T$ in
$^{120}$Sn and $^{148}$Au can be understood in terms of simple 
shape effects caused by pairing correlations.  Fluctuations in
pairing field lead to a slowly vanishing pairing gap [Fig. \ref{Sn120}
(a)], which influences the
structural properties even at moderate $T$ ($\sim$1 MeV).  We
found that the low-$T$ structure and hence the GDR width are 
quite sensitive to the change of the pairing field [Fig. \ref{Sn120}
(b)].  

\section{Conclusions}
It has been shown in the present lecture that the MHFB and MBCS
theories are microscopic approaches, which take into account thermal fluctuations of 
quasiparticle number. These large thermal fluctuations smooth out the sharp SN phase transition 
in finite nuclei. As a result, the thermal pairing gap does not
collapse, but decreases monotonously with increasing temperature $T$, 
remaining finite even at $T$ as high as 4 - 5 MeV. This non-vanishing
thermal pairing gap keeps the width of GDR remain
almost constant at low $T$ ($\leq$ 1 MeV for $^{120}$Sn) when it is
included in the PDM. In this way the PDM becomes a semi-microscopic
model that is able to describe the temperature dependence of the
GDR width in a consistent way within a large temperature interval
starting from very low $T$, where the GDR width is nearly
$T$-independent, to the region when the width increases with $T$ (1
$\leq T\leq$ 3 - 4 MeV), and up to the region of high $T$ ($T>$ 4 - 5 
MeV), where the width seems to saturate in tin isotopes.

\end{document}